# Higher Order Nonlinear Dynamics in AlGaAs Doped Glass Photonic Crystal Fibers at Sub Pico-Joule Energy

**Mohit Sharma[1, 2], D. Vigneswaran[3], Julia S. Skibina[4], Vinoth Kumar*[, 5] and S. Konar[1]**

[1]*Department of Physics, Birla Institute of Technology, Mesra, India*
[2]*Department of Physics, PDM University, Bahdurgarh, India*
[3]*Department of Physics, University College of Engineering, Anna University, Ramanathapuram, India*
[4]*SPE LLC "Nonstructural Glass Technology", Saratov, Russia*
[5]*Department of Electronics and Electrical Engineering, Karunya Institute of Science and Technology, Coimbatore, India*
*Corresponding author: kvinoth_kumar84@yahoo.in*

**Abstract:** - In the present paper, a unique semiconductor doped glass photonic crystal fiber has been designed which is suitable for soliton propagation at sub pico-Joule energy. The fiber promises to yield low and uniform anomalous dispersion profile and very large optical nonlinearities $25211\ W^{-1}km^{-1}$ at telecommunication wavelength, thus facilitating soliton formation at ultralow energy. The observed magnitude of nonlinearity is the highest reported ever till date in AlGaAs doped glass. The soliton dynamics has been investigated taking into account of higher-order dispersions and nonlinearities. Propagating solitons breadth and experience large frequency shift, which decreases with the increase in the initial pulse width and increases with the increase in pulse energy. Temporal width of solitons oscillate, the frequency and amplitude of oscillations increase with the increase in the value of pulse energy.

## 1. Introduction

Optical solitons in photonic crystal fibers (PCF) open up new opportunities, which find applications in telecommunications, sensing technologies, supercontinuum and bio-photonics [1-10]. Among notable efforts on this topic, Luan et al. [11], Ouzounov et al. [12] and Saleh et al. [13] have investigated soliton propagation in hollow-core photonic band gap fibers. On the other hand, Wadsworth et al. [14], Khan et al. [15] and Nishizawa et al. [16] have investigated soliton propagation in solid core PCFs. Till date, though most of the investigations on optical soliton propagation have been carried out in undoped fibers and undoped optical waveguides, in recent days significant attention has been also paid to investigate optical soliton propagation in doped fibers as well as waveguides made of chalcogenide glass, and semiconductors such as silicon and AlGaAs [19-22]. Large refractive index of semiconductor waveguides can lead to extremely tightly confined optical modes (can go to sub μm), that can produce optical nonlinearities exceeding $3 \times 10^5\ W^{-1}km^{-1}$ [20-22]. Such large nonlinearities eventually allow nonlinear optics to become active at sub-Watt power level. In view of this, several authors have investigated soliton generation in praseodymium fluoride fiber laser, erbium doped fiber ring laser and erbium doped fiber waveguides [23-25]. Recently, Barviau et al. [18] have investigated the enhanced soliton self-frequency shift and continuous wave supercontinuum generation in $GeO_2$ doped index guiding PCFs. Hickmann et al. [26] have investigated the effects of fifth order nonlinearity on optical pulse propagation in semiconductor doped glass fibers. In addition, efficient supercontinuum generation in a compact waveguide composed of doped high index glass has been also reported recently [27].

Recently nonlinear properties of AlGaAs in different waveguides have received significant attention [28-36]. For instance, Aster *et al.* [32] have reported wavelength conversion of 10 Gb/s signal by XPM in 4.5 mm long GaAs-AlGaAs waveguides. Aitchison *et* *al.* [33] have experimentally reported the values of nonlinear optical coefficients of AlGaAs in the half-band-gap spectral region. They have studied the polarization effect on XPM and SPM from band structure calculation. El-Ganainy *et al.* [34] have demonstrated the method of inducing optical Kerr-effect to generate high stable soliton pulses in dispersion inverted highly nonlinear AlGaAs nanowires and generated self-localized solitons using very low power. Siviloglou *et al.* [35] have investigated the enhanced third-order nonlinear effects in optical AlGaAs nanowires. Moreover, Kao *et al.* [36] have studied the Raman effect in AlGaAs waveguide numerically and experimentally. They have reported the depolarized Raman gain spectra upto 300 cm$^{-1}$ in $Al_{0.24}Ga_{0.76}As$ at the pump wavelengths of 515 and 1550 nm.

In view of above development, it would be worth examining the propagation characteristics of optical solitons in an AlGaAs doped PCF. In the doped fiber, appropriate design shall yield anomalous dispersion with low magnitude at the chosen wavelength. Therefore, the combination of extremely large nonlinearity and low anomalous dispersion shall inevitably make propagation dynamics of optical solitons interesting in AlGaAs doped PCFs. The organization of the paper is as follows: In section 2, we have designed and studied the optical properties of the fiber. Soliton dynamics has been investigated in section 3. A brief conclusion has been added in section 4.

## 2. AlGaAs doped PCF and its optical properties

Several The primary objective of the present work is to design a semiconductor doped PCF that promises high nonlinearity and low anomalous dispersion at the telecommunication wavelength. In order to achieve this, we assume that the silica fiber is doped with AlGaAs which will ensure large nonlinearity. Furthermore, the PCF cladding structure is designed in such a manner that it promises low anomalous dispersion over a wide range of operating wavelength. The motivation of doping AlGaAs has been explained in next paragraph. The cross section of the proposed PCF structure has been illustrated in Fig.1(a). The cladding region is composed of seven rings of small as well as large air holes. Each ring of air holes is hexagonal and the diameter $d_i$ of all air holes in the i$^{th}$ ring is same. However, the size of air holes in two adjacent rings is not necessarily same. Air holes of the first two rings are of same size ($d_1 = d_2 = 0.5\ \Lambda$, $\Lambda$ the hole pitch). Air holes of third and fourth rings ($d_3 = d_4 = 0.7\ \Lambda$) are slightly larger in comparison to the air holes in the first two rings. Air holes in the three outermost rings are largest in size ($d_5 = d_6 = d_7 = 0.9\ \Lambda$). First two rings of air holes ensure tight confinement of the fundamental mode. This unique three layer structure helps to achieve low anomalous dispersion and single mode operation over a large range of wavelengths. On the other hand, a fiber in which all air holes are of same size exhibits large dispersion, consequently requires large power for creation of optical solitons. A typical optical mode in the designed fiber has been demonstrated in Fig.1(b) which is calculated at telecommunication wavelength.

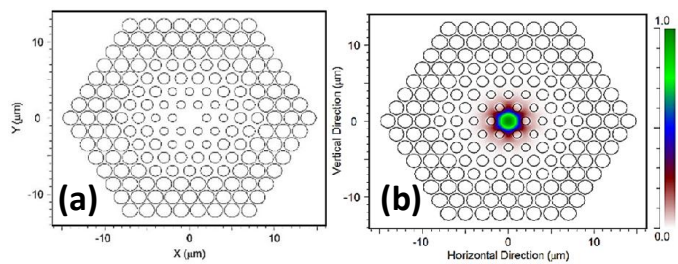

Figure 1 (a) Cross section of the designed PCF, (b) transverse mode profile of the proposed fiber at 1550 nm. $d_1 = d_2 = 0.5 \times \Lambda\ \mu m$, $d_3 = d_4 = 0.7 \times \Lambda\ \mu m$, $d_5 = d_6 = d_7 = 0.9 \times \Lambda\ \mu m$ and $\Lambda = 2\ \mu m$.

As usual, the optical properties of the doped PCF has been investigated employing FDTD method [27-30]. Fig.2 demonstrates the variations of the effective refractive index $n_{eff}$ and the total chromatic dispersion for different hole pitch $\Lambda$. At low values of $\Lambda$, the fiber dispersion is in the normal dispersion regime, with the increase in the value of Λ, the magnitude of dispersion decreases, finally above certain value of Λ (i.e., 1.8 μm), the fiber becomes anomalously dispersive at telecommunication wavelength. Note that the fiber with $\Lambda = 1.9\ \mu m$ exhibits low and uniform anomalous dispersion over a wide range of operating wavelengths. The magnitude of the dispersion for $\Lambda = 1.9\ \mu m$ at 1550 nm wavelength is ~ 44 ps/km/nm, which is appropriate for soliton generation at telecommunication wavelength. Fig.3 demonstrates the variation of the mode field area of the fundamental mode of the doped fiber for different hole pitch Λ. From figure it is evident that the effective mode field area increases with the increase in the wavelength. In a photonic crystal fiber this behavior is usual since the fiber remains single mode over a large range of wavelengths and that is possible only when area of the mode field increases with the increase in wavelength.

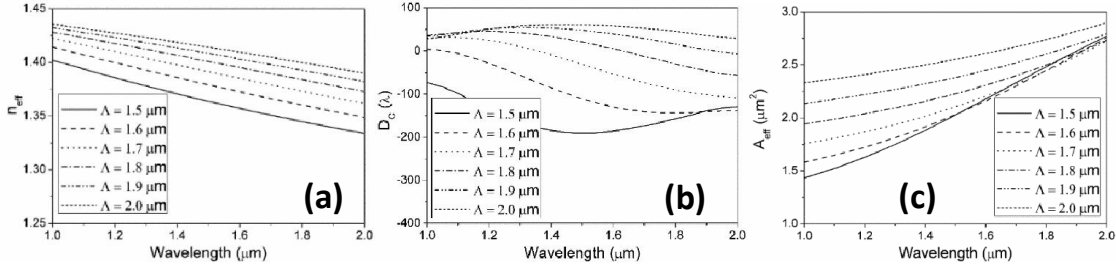

Figure 2 (a) Variation of effective refractive index, (b) dispersion profiles and (c) effective mode area of the fiber at different air hole pitch with $d_1 = d_2 = 0.5 \times \Lambda\ \mu m, d_3 = d_4 = 0.7 \times \Lambda\ \mu m, d_5 = d_6 = d_7 = 0.9 \times \Lambda\ \mu m$.

In general the effective nonlinearity in a photonic crystal fiber is described by $\gamma = (2\pi n_2/\lambda A_{eff}) \times 10^3\ W^{-1}km^{-1}$, where $n_2$ is the nonlinear coefficient associated with third order susceptibility $\chi^{(3)}$ (i.e., Kerr nonlinearity), and $A_{eff}$ is the effective mode area. At a given wavelength, the effective nonlinearity is inversely proportional to effective mode area and directly proportional to nonlinear co-efficient ($n_2$). Therefore, the effective nonlinearity in the fiber can be enhanced either by reducing $A_{eff}$ or by choosing a material with large $n_2$. To achieve large nonlinearity, another alternative is to dope glass fiber with a semiconducting material. However, in this case, not only the third order susceptibility, another term i.e., fifth order optical susceptibility $\chi^{(5)}$ is also important. In silica fiber, since the value of $|\chi^{(5)}|$ is negligible in comparison to $|\chi^{(3)}|$, only the third order nonlinearity is generally taken into account to describe optical pulse propagation in silica PCFs. However, due to relatively large value of $\chi^{(5)}$, in semiconductor doped glass PCFs, one needs to incorporate fifth order nonlinear term also to describe optical soliton propagation. Therefore, we assume that the PCF core is doped with AlGaAs and proceed to estimate third as well as fifth order nonlinearities. Our choice of AlGaAs doping is motivated by the fact that it possesses a very large third ($\chi^{(3)}$) and fifth ($\chi^{(5)}$) order susceptibilities in comparison to silica. This necessitated evaluation of two effective nonlinear co-efficients of the designed fiber, one due to the usual third order nonlinear coefficient $\gamma_{AlGaAs} = (2\pi n_2/\lambda A_{eff}) \times 10^3\ W^{-1}km^{-1}$, the other one $\gamma'_{AlGaAs}$ is due to the fifth order susceptibility $\chi^{(5)}$, which is defined as $\gamma'_{AlGaAs} = (2\pi n_4/\lambda A_{eff}^2) \times 10^3\ W^{-2}km^{-4}$ [42], where $n_2 = \frac{3\chi^{(3)}}{8n_0}$ and $n_4 = \frac{5\chi^{(5)}}{16n_0}$, $n_0$ is the linear refractive index. For AlGaAs, $n_2 = 1.5 \times 10^{-13}$ cm$^2$/W [36] and $n_4 = -5 \times 10^{-23}$ cm$^4$/W$^2$ [36]. With above values of $n_2$ and $n_4$, we have evaluated $\gamma_{AlGaAs}$ and $\gamma'_{AlGaAs}$. In addition, we have also evaluated $\gamma_{Si}$ for undoped silica fiber with same structure. Due to small effective area of the fundamental mode, the PCF exhibits very high nonlinear coefficient. The values of the third ($\gamma_{AlGaAs}$) and fifth ($\gamma'_{AlGaAs}$) order nonlinear coefficient of AlGaAs doped PCF at 1550 nm for $\Lambda = 1.9\ \mu m$ are 25211 $W^{-1}km^{-1}$ and $-348\ W^{-2}km^{-4}$, respectively, and for silica PCF $\gamma_{Si}$ and $\gamma'_{Si}$ are 36 $W^{-1}km^{-1}$ and $2.4 \times 10^{-13}\ W^{-2}km^{-4}$, respectively. The variations $\gamma_{Si}$ and $\gamma_{AlGaAs}$ have been demonstrated in Fig.4. Due to large $n_2$ of AlGaAs, $\gamma_{AlGaAs}$ at telecommunication wavelength 1.55 $\mu m$ is four order larger than $\gamma_{Si}$. The value of $\gamma_{AlGaAs}$ decreases with the increase in wavelength, this is obvious since the modefield size increases with increasing wavelength, consequently decreasing effective nonlinearity. The variation of the effective fifth order nonlinearity $\gamma'_{AlGaAs}$ with wavelength has been demonstrated in Fig.5. As expected, the magnitude of the effective fifth order nonlinearity decreases with the increase in wavelength.

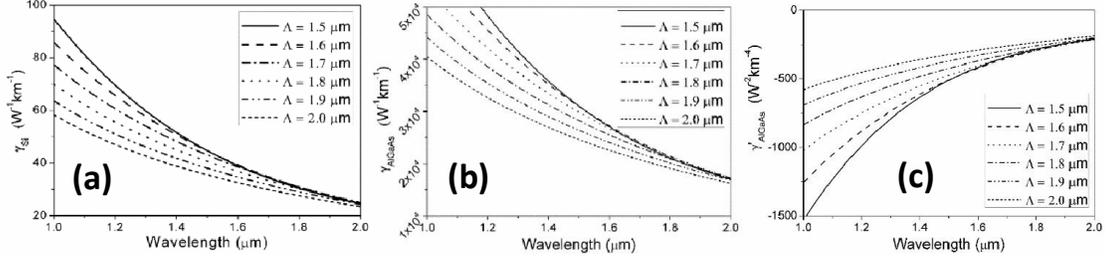

Figure 4 Variation of nonlinear co-efficients of undoped and semiconductor doped silica PCF designed in this section. (a) The undoped silica fiber. (b) The semiconductor doped fiber and (c) fifth order susceptibility $\chi^{(5)}$.

In order to examine the single modeness of the fiber, we have demonstrated the variation of effective V-parameter with wavelength in Fig.6. The value of $V_{eff}$ gradually decreases with the increase in the value of wavelength. From the figure it is evident that $V_{eff} \leq 4.1$, hence, the fiber is endlessly single mode over a wide range of wavelengths [20].

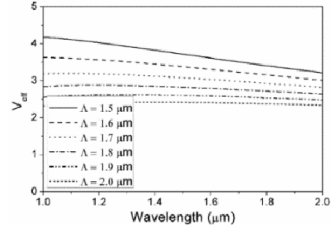

Figure 6 Effective V-parameter of the fiber at different air hole pitch with $d_1 = d_2 = 0.5 \times \Lambda\ \mu m, d_3 = d_4 = 0.7 \times \Lambda\ \mu m, d_5 = d_6 = d_7 = 0.9 \times \Lambda\ \mu m$.

## 3. Modified nonlinear Schrödinger Equation

In this section, we investigate optical soliton propagation in the PCF which has been designed in the previous section. Optical pulse propagation through this fiber can be described by the well-known generalized nonlinear Schrödinger equation (GNLSE) [43], which can be written as:

$$i\frac{\partial}{\partial z}A(z,\tau) - \frac{1}{2}\beta_2\frac{\partial^2}{\partial \tau^2}A(z,\tau) - \frac{i}{6}\beta_3\frac{\partial^3}{\partial \tau^3}A(z,\tau) + \frac{1}{24}\beta_4\frac{\partial^4}{\partial \tau^4}A(z,\tau) + \gamma_{AlGaAs}|A|^2 A + \gamma'_{AlGaAs}|A|^4 A + i\frac{\gamma_{AlGaAs}}{\omega_0}\frac{\partial}{\partial \tau}(|A|^2 A) - \gamma_{AlGaAs}T_R A\frac{\partial |A|^2}{\partial \tau} = 0, \quad (1)$$

where $A(z,\tau)$ is the envelope of the electric field of the propagating pulse, $\beta_n = \frac{d^n\beta}{d\omega^n}$ is the nth order dispersion at pumping wavelength $\lambda_P$; in above equation, we have kept up to fourth order of dispersion terms. In writing above equation, we have neglected terms which are arising due to the nonlinearity of glass, since as elucidated earlier, these are few order smaller in comparison to the corresponding term associated with AlGaAs. $T_R$ is the first moment of Raman response function R which can be written in the form $T_R = \int_0^\infty \tau R(\tau)d\tau$. R is the nonlinear response function, which consists of an instantaneous electronic response and a contribution from delayed Raman response. The Raman response function $R(\tau)$ may be defined as $R(\tau) = (1-f_r)\delta(\tau) + f_r h_r(\tau)$, where $f_r$ is the Raman fraction and $h_r(\tau) = \frac{\tau_1^2+\tau_2^2}{\tau_1\tau_2^2}exp\left(\frac{-\tau}{\tau_2}\right)sin\left(\frac{\tau}{\tau_1}\right)$ is the Raman response function; $\tau_1 = \frac{1}{\Omega_R}$ and $\tau_2$ is the damping time of vibration, $\Omega_R$ is the vibrational frequency. For silica, $f_r = 0.18$, $\tau_1 = 12.5$ fs, $\tau_2 = 32$ fs and $T_R = 2.5$ fs [20]. For AlGaAs, the Raman fraction $f_r$ and Raman response function $h_r(\tau)$ can be calculated adopting the procedure outlined in Zhang *et al.* [44]. The numerically evaluated value of Raman response function of AlGaAs by using Kramers-Kronig relation [44] has been demonstrated in Fig.7. The corresponding calculated values of different parameters associated with AlGaAs are: $f_r = 0.047$, $\tau_1 = 3.6$ fs, $\tau_2 = 2.8$ fs and $T_R = 5$ fs.

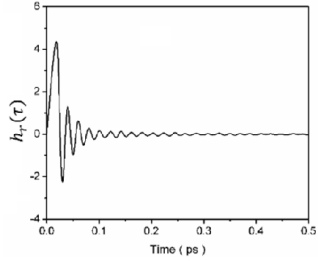

Figure 7 Calculated Raman response function of the AlGaAs.

The modified generalized nonlinear Schrödinger Equation (GNLSE) can be rewritten in the following form

$$i\frac{\partial A}{\partial z} - \frac{\beta_2}{2}\frac{\partial^2 A}{\partial \tau^2} - \frac{i}{6}\beta_3\frac{\partial^3 A}{\partial \tau^3} + \frac{1}{24}\beta_4\frac{\partial^4 A}{\partial \tau^4} + \gamma_{AlGaAs}|A|^2 A + \gamma'_{AlGaAs}|A|^4 A = iG, \quad (2)$$

where, $G = -\frac{\gamma_{AlGaAs}}{\omega_0}\frac{\partial}{\partial \tau}(|A|^2 A) - i\gamma_{AlGaAs}T_R A\frac{\partial |A|^2}{\partial \tau}$. We note that in absence of the right hand side term, equation (2) does possess a Lagrangian density. However, the right hand side term G is small in comparison to other terms in equation (2), hence, this equation can be solved treating G as a perturbation. In the absence of right hand side, equation (2) can be solved analytically using variational formulation [45-47]. In order to solve the unperturbed equation, we identify a Lagrangian density $\Gamma$ such that variation of this density satisfies $\delta \int_{-\infty}^{\infty}\int_{-\infty}^{\infty}\Gamma d\tau dz = 0$, where, the Lagrangian density $\Gamma$ is given by

$$\Gamma = i\left(A\frac{\partial A^*}{\partial z} - A^*\frac{\partial A}{\partial z}\right) - \frac{\beta_2}{2}\left|\frac{\partial A}{\partial \tau}\right|^2 - \frac{i}{12}\beta_3\left(\frac{\partial^2 A}{\partial \tau^2}\frac{\partial A^*}{\partial \tau} - \frac{\partial^2 A^*}{\partial \tau^2}\frac{\partial A}{\partial \tau}\right) + \frac{1}{72}\beta_4\left(\frac{\partial^3 A}{\partial \tau^3}\frac{\partial A^*}{\partial \tau} - \frac{\partial^3 A^*}{\partial \tau^3}\frac{\partial A}{\partial \tau}\right) + \frac{1}{2}\gamma_{AlGaAs}|A|^4 + \frac{1}{3}\gamma'_{AlGaAs}|A|^6.$$

(4) The unperturbed Schrödinger equation can be recovered using

$$\frac{\partial}{\partial z}\left(\frac{\partial \Gamma}{\partial\left(\frac{\partial A}{\partial z}\right)}\right) + \frac{\partial}{\partial \tau}\left(\frac{\partial \Gamma}{\partial\left(\frac{\partial A}{\partial \tau}\right)}\right) - \frac{\partial \Gamma}{\partial A} = 0, \quad \text{5(a)}$$

$$\frac{\partial}{\partial z}\left(\frac{\partial \Gamma}{\partial\left(\frac{\partial A^*}{\partial z}\right)}\right) + \frac{\partial}{\partial \tau}\left(\frac{\partial \Gamma}{\partial\left(\frac{\partial A^*}{\partial \tau}\right)}\right) - \frac{\partial \Gamma}{\partial A^*} = 0. \quad \text{5(b)}$$

In order to analytically investigate the optical pulse dynamics in the semiconductor doped PCF, we take a simple Gaussian ansatz as follows: $A(Z,\tau) = \sqrt{\frac{E_0}{\sqrt{\pi}T(Z)}}\exp\left(-(1+iC(Z)\frac{(\tau-t_p(Z))^2}{2T(Z)^2} - i\Omega(\tau - t_p(Z)) + \phi(Z)\right)$, where, $E_0$ is the normalized energy of the pulse, $T$ is the pulse width, $t_p$ is the position of the pulse center, $C$ is the chirp, $\Omega$ is the nonlinear frequency shift and $\phi$ is the phase of the pulse. The peak amplitude $p(Z)$ of the pulse may be defined as $p(Z) = \sqrt{\frac{E_0}{\sqrt{\pi}T(Z)}}$. We now evaluate a finite dimensional reduced Lagrangian $L\left(T, t_p, C, \Omega, \phi, \frac{\partial T}{\partial z}, \frac{\partial t_p}{\partial z}, \frac{\partial C}{\partial z}, \frac{\partial \Omega}{\partial z}, \frac{\partial \phi}{\partial z}\right)$ such that the evolution equations of the pulse parameters are obtainable from $\delta \int_{-\infty}^{\infty} L\, dZ = 0$, where the reduced Lagrangian $L$ is given by $L = \int_{-\infty}^{\infty}\Gamma d\tau$,

$$= 2E_0\left(-\frac{1}{4}\frac{\partial C}{\partial z} + \frac{C}{2T}\frac{\partial T}{\partial z} + \Omega\frac{\partial t_p}{\partial z} + \frac{\partial \phi}{\partial z}\right) + \frac{\beta_2 E_0}{2T^2}\left(\frac{1+C^2}{2} + \Omega^2 T^2\right) - \frac{\beta_3 \Omega E_0}{12 T^2}(3 + 3C^2 + 2\Omega^2 T^2) - \frac{\beta_4 \Omega E_0}{48 T^3}(1 + C^2 + 2\Omega^2 T^2) - \frac{E_0^2}{2\sqrt{2\pi}T}\gamma_{AlGaAs} + \frac{E_0^3}{3\sqrt{3\pi}T}\gamma'_{AlGaAs}. \quad (7)$$

The dynamical equations of different pulse parameters can be easily obtained using Euler-Lagrange Equation, $\frac{d}{dz}\left(\frac{\partial \langle L\rangle}{\partial\left(\frac{\partial r_j}{\partial z}\right)}\right) - \frac{\partial \langle L\rangle}{\partial r_j} = 0$, where, $r_j = T, C, t_p, \Omega$ and $\phi$. We now introduce the perturbed vibrational procedure to solve equation (2) with small but finite value of G. Since the GNLSE contains perturbation terms, therefore, the modified Euler-Lagrange equations can be written as,

$\frac{d}{dz}\left(\frac{\partial \langle L \rangle}{\partial \left(\frac{\partial r_j}{\partial z}\right)}\right) - \frac{\partial \langle L \rangle}{\partial r_j} = i \int_{-\infty}^{\infty}\left(R\left(\frac{\partial A^*}{\partial r_j}\right) - R^*\left(\frac{\partial A}{\partial r_j}\right)\right)d\tau.$ On inserting the value of L from equation (7) into above Euler-Lagrange equation, we obtain a set of first order ordinary differential equations obeyed by the pulse parameters, which are as follows,

$$\frac{\partial T}{\partial z} = \frac{\beta_2 C}{2T} + \frac{\beta_3 \Omega C}{2T} - \frac{\beta_4 \Omega C}{24T^2}, \quad (10)$$

$$\frac{\partial C}{\partial z} = \frac{\beta_2(1+C^2)}{2T^2} + \frac{\beta_3 \Omega(1+C^2)}{2T^2} - \frac{\beta_4 \Omega(3+3C^2-2\Omega^2 T^2)}{48T^3} - \frac{E_0 \gamma_{AlGaAs}}{2\sqrt{2\pi}T} + \frac{E_0^2 \gamma'_{AlGaAs}}{3\sqrt{3\pi}T} - \frac{E_0 \Omega \gamma_{AlGaAs}}{2\sqrt{2\pi}T\omega_0}, \quad (11)$$

$$\frac{\partial t_p}{\partial z} = \frac{\beta_2 \Omega}{2} - \frac{\beta_3(1+C^2+2\Omega^2 T^2)}{8T^2} + \frac{\beta_4(1+C^2+6\Omega^2 T^2)}{36T^2} - \frac{3E_0 \gamma_{AlGaAs}}{4\omega_0 \sqrt{2\pi}T}, \quad (12)$$

$$\frac{\partial \Omega}{\partial z} = \frac{E_0}{2\sqrt{2\pi}T^3}\left(\frac{\gamma_{AlGaAs}}{\omega_0}C - \gamma_{AlGaAs}T_R\right). \quad (13)$$

In addition to above, the pulse amplitude $p(Z)$ and the pulse width T is related to the normalized pulse energy $E_0$ through $E_0 = \sqrt{\pi}p^2 T$, which is the well-known energy conservation law. Equations (10) – (13) describe the nonlinear propagation of optical solitons as evolutions of pulse parameters with distance.

To examine the pulse evolution through the PCF, the set of differential equations (10)-(13) need to be solved numerically. Note that though the third equation depends on the pulse width T, Chirp C and nonlinear frequency shift $\Omega$, dynamical equations of T, C, and $\Omega$ do not depend on the position of pulse center $t_p$. We now consider the propagation of pulses of 50 fs duration at 1550 nm wavelength through the PCF. The relevant parameters of the PCF at this wavelength are: $\beta_2 = -5.61\ ps^2/km$, $\beta_3 = 4.03 \times 10^{-3}\ ps^3/km$, $\beta_4 = 4.86 \times 10^{-5}\ ps^4/km$, $\gamma_{AlGaAs} = 25211\ W^{-1}km^{-1}$ and $\gamma'_{AlGaAs} = -348\ W^{-2}km^{-4}$.

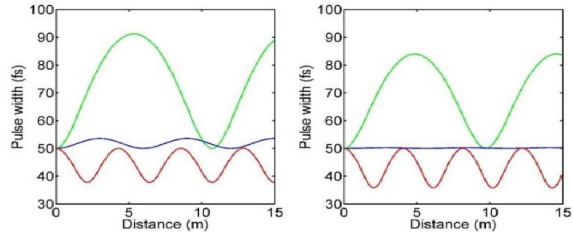

Figure 8 Variation of pulse width along the length of the PCF. In the left panel $\gamma'_{AlGaAs} = 0$, and $\gamma'_{AlGaAs} \neq 0$ in the right panel. Green line $E_0 = 0.1\ fJ$, blue line $E_0 = 0.125\ fJ$ and red line $E_0 = 0.15\ fJ$.

The variation of the temporal duration i.e., pulse width has been depicted in Fig.8 at three different pulse energies. The quintic or fifth order nonlinearity is absent in the left panel, while this is nonzero in the right hand panel. At 0.1 *fJ* pulse energy, the pulse first broaden, reaches to a large value then compresses till the initial value is attained. This process then repeats, and occurs for both $\gamma'_{AlGaAs} = 0$, and $\gamma'_{AlGaAs} \neq 0$. Initially the pulse duration increases since the available pulse energy is less than that is required for soliton formation. At 0.125 *fJ* energy, the pulse propagates almost preserving its pulse duration. At higher power, i.e., at 0.15 *fJ* or above, the pulse gets compressed first and then expands, again compresses and expands. This process repeats, while the pulse propagates through the fiber. However, at 0.15 *fJ* energy, the pulse width never exceeds its initial width. The variations of the pulse width for $\gamma'_{AlGaAs} = 0$, and $\gamma'_{AlGaAs} \neq 0$ are identical, except, the amplitude of pulse oscillation and frequency of breathing are marginally different.

We now investigate the evolutions of nonlinear frequency shift, pulse chirp and the location of the pulse center. In addition to the variation of pulse width, Fig.9 demonstrates the variations of above parameters. As the pulse propagates, the frequency shift increases with the increasing distance. The pulse chirp oscillates as the pulse propagates leading to the oscillation of pulse width. The pulse center shifts since the frequency of the pulse reduces due to self-frequency shift.

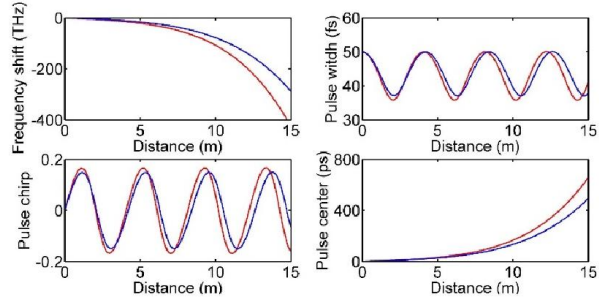

Figure 9 Variations of frequency shift, pulse chirp, pulse width, and shift of pulse center with propagation distance. Initial pulse width 50 fs, pulse energy 0.15 fJ. The red line is for $\gamma'_{AlGaAs} \neq 0$, and blue line $\gamma'_{AlGaAs} = 0$.

In order to check the effect of pulse duration on the pulse dynamics, we have extended our investigation for three different initial pulse widths. The variations of pulse parameters along the propagation direction have been depicted in Fig.10. The frequency shift decreases with the increase in the value of pulse duration. Note that at a fixed propagation distance, the frequency shift is augmented due to fifth order nonlinearity. From figure it is evident that with the increase in the value of pulse duration, the pulse center moves slowly. The role of the fifth order nonlinearity is to make the shift of the pulse center marginally faster. The variation of the pulse chirp is oscillatory in nature, though it is not evident at a short propagation distance. The oscillation frequency decreases with the increase in the value of pulse duration. The chirp in the pulse leads to the variation of the pulse width, and it is amply clear from the figure that with the increase in the value of initial pulse width, the variation of pulse width with distance of propagation becomes slower.

Before closing, it would be worth examining the effect of pulse energy on the pulse dynamics. In order to do that we have demonstrated the variations of frequency shift and relevant pulse parameters of a 50 fs pulse in Fig.11. The magnitude of the nonlinear frequency shift, as evident from figure, increases with the increase in the pulse energy. The role of fifth order nonlinearity is to augment frequency shift. With the increase in the pulse energy, the shift in the location of pulse center becomes faster. The role of fifth order nonlinearity is to make this change faster. The pulse chirp and pulse width oscillate as the pulse propagate along the PCF. The frequency and amplitude of oscillations, though not quantified, increase with the increase in the value of pulse energy.

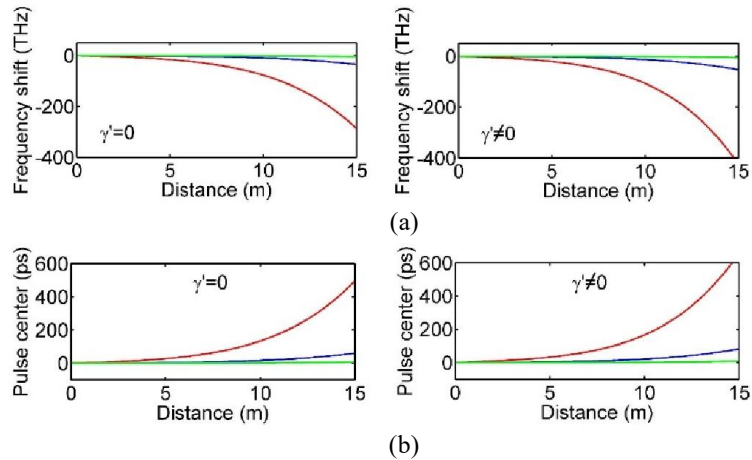

(a)

(b)

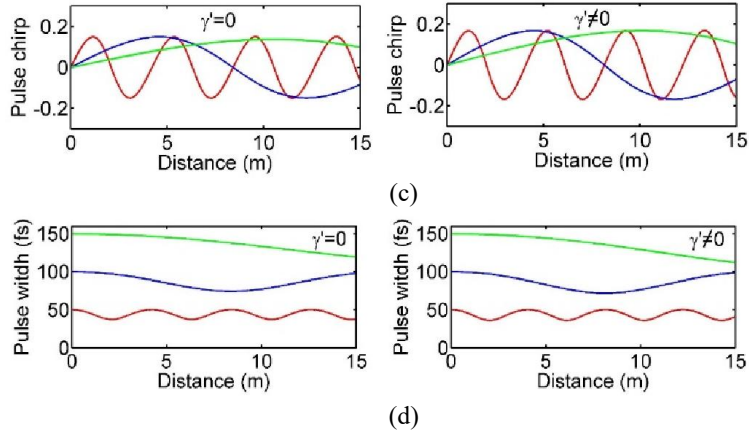

(c)

(d)

Figure 10 Evolution of the pulse at different pulse duration; (a) frequency shift, (b) pulse center, (c) pulse chirp and (d) pulse width. Left panel in absence of fifth order nonlinearity ($\gamma'_{AlGaAs} = 0$), the right panel in presence of fifth order nonlinearity ($\gamma'_{AlGaAs} \neq 0$). Red line for 50 fs, blue line for 100 fs and green line for 150 fs pulses. Pulse energy 0.15 fJ.

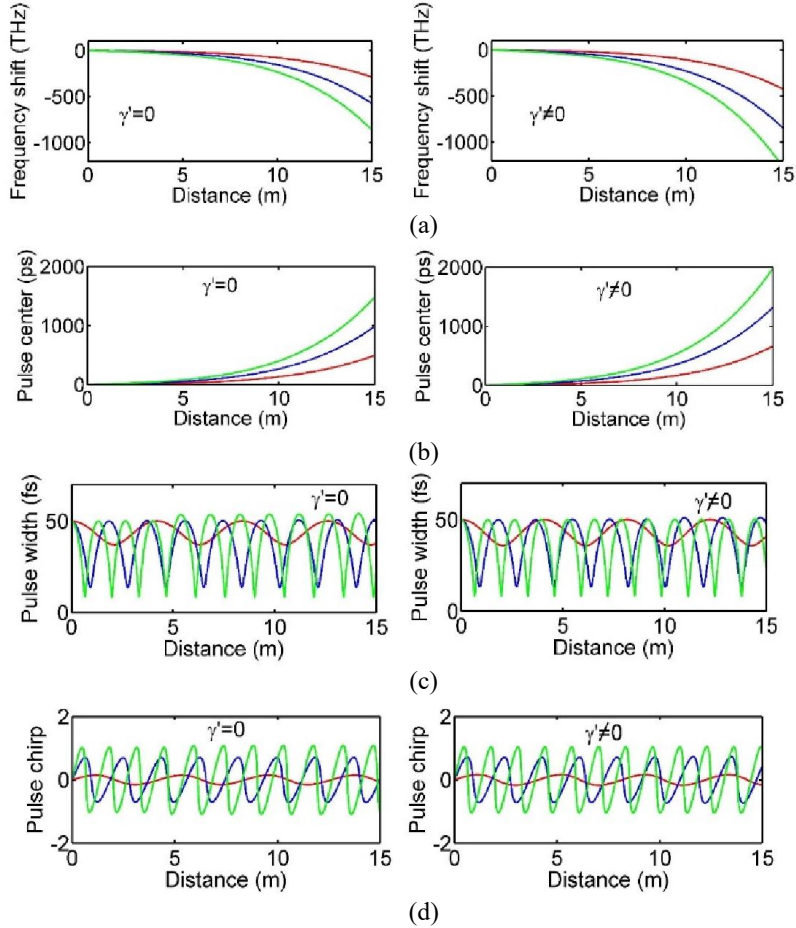

(a)

(b)

(c)

(d)

Figure 11 Evolution of the pulse at different pulse energy; (a) frequency shift, (b) pulse center, (c) pulse width and (d) pulse chirp. Red, blue and green line represents 0.15 fJ, 0.3 fJ and 0.45 fJ, respectively.

## 4. Conclusion

In conclusion, we have designed a semiconductor (AlGaAs) doped PCF with seven rings of air holes. These rings are arranged in an increasing order of air hole diameter starting from the core. The proposed PCF promises to yield flat anomalous dispersion profile and large nonlinear coefficient over a wide range of operating wavelengths. The values of the third ($\gamma_{AlGaAs}$) and fifth ($\gamma'_{AlGaAs}$) order nonlinear coefficients of *AlGaAs* at 1550 nm wavelength are 25211 $W^{-1}km^{-1}$ and $-348$ $W^{-2}km^{-4}$, respectively. We have investigated the dynamics of soliton propagation inside the designed fiber taking into account of the effect of fourth order dispersion, third and fifth order nonlinearities, intra-pulse Raman scattering and self-steepening. We have solved the modified nonlinear Schrödinger equation using variational method and obtained the several ordinary differential equations for various pulse parameters. It is realised that the designed fiber is suitable for propagation of optical solitons at ultralow energy owing to extremely large nonlinearity and very low group velocity dispersion. While propagating, optical solitons experiences large frequency shift, and pulse width breadths. The frequency shift decreases with the increase in initial pulse width. The presence of finite fifth order nonlinearity increases the frequency shift. We have examined the role of pulse energy in determining the pulse dynamics. The nonlinear frequency shift increases with the increase in the value of pulse energy. The pulse chirp and pulse width oscillate as the pulse propagates along the PCF. The frequency and amplitude of oscillations, though not quantified, increase with the increase in the value of pulse energy.